\documentclass[conference]{IEEEtran}
\IEEEoverridecommandlockouts
\usepackage{cite}
\usepackage{amsmath,amssymb,amsfonts}
\usepackage{algorithmic}
\usepackage{graphicx}
\usepackage{textcomp}
\usepackage{xcolor}
\usepackage{todonotes}
\usepackage{lipsum,graphicx,subcaption}
\usepackage{multirow}
\usepackage{array}
\def\BibTeX{{\rm B\kern-.05em{\sc i\kern-.025em b}\kern-.08em
    T\kern-.1667em\lower.7ex\hbox{E}\kern-.125emX}}
\begin{document}
\title{Towards Benchmarking Power-Performance Characteristics of Federated Learning Clients \\

}

\author{\IEEEauthorblockN{Pratik Agrawal, Philipp Wiesner, and Odej Kao}
\IEEEauthorblockA{
\textit{Technische Universität Berlin}\\
\{pratikkumar.vijaykumar.agrawal, wiesner, odej.kao\}@tu-berlin.de}
}

\maketitle

\begin{abstract}
Federated Learning (FL) is a decentralized machine learning approach where local models are trained on distributed clients, allowing privacy-preserving collaboration by sharing model updates instead of raw data.
However, the added communication overhead and increased training time caused by heterogenous data distributions results in higher energy consumption and carbon emissions for achieving similar model performance than traditional machine learning.
At the same time, efficient usage of available energy is an important requirement for battery-constrained devices.
Because of this, many different approaches on energy-efficient and carbon-efficient FL scheduling and client selection have been published in recent years.

However, most of this research oversimplifies power-performance characteristics of clients by assuming that they always require the same amount of energy per processed sample throughout training.
This overlooks real-world effects arising from operating devices under different power modes or the side effects of running other workloads in parallel.
In this work, we take a first look on the impact of such factors and discuss how better power-per\-for\-mance estimates can improve energy-efficient and carbon-efficient FL scheduling. 
  

\end{abstract}


\section{Introduction}

While FL\cite{FL_basepaper} mitigates privacy and data transfer overhead issues associated with centralized ML training, it has different set of challenges. 
Studies\cite{fl_power_cost} have shown that FL consumes more energy and emits significantly more carbon to reach the same model performance as centrally trained ML models.  

The usage of Internet of Things (IoT) and edge devices to train machine learning models in distributed and federated learning settings regularly have further worsened the energy and environment implications of AI/ML training.
 Moreover, regulators responsible for AI policy also explicitly put emphasis on sustainability and environmental aspects of AI\footnote{https://digital-strategy.ec.europa.eu/en/library/ethics-guidelines-trustworthy-ai}. 
 Additionally,  edge devices are battery powered and operate under strict energy budgets. 
 To manage efficient usage of this limited battery power for both the non-FL baseloads and the FL training, it is necessary to profile power-performance characteristics of FL training under realistic scenarios. 

To improve efficiency, researchers have proposed energy-efficient\cite{zhou2022joint,zaw2021energy,kim2021autofl} and carbon-efficient\cite{wiesner2023fedzero, guler2021framework} approaches for FL scheduling and client selection.
 FL has also seen vast amount of work in benchmarking\cite{rupprecht2022performance} in recent years.
However, these benchmarks are often missing the critical energy consumption metrics of FL and are predominantly explored in server based simulated environments. 
Moreover, the benchmarks that include energy consumption metrics are missing granular level energy consumption data under the real-world impact factors such as non-FL baseloads and power modes of the devices. 
These approaches\cite{wiesner2023fedzero} rely on analytical energy models that are based on metrics such as CPU compute cycles or Floating Point Operations Per Second (FLOPS) to calculate energy required per batch. 
These models do not accurately take into account factors such as power modes and baseloads of the underlying FL device. 
\section{Problem Statement}

Current FL studies over simplify the power consumption characteristics of FL clients. 
In comparison to simulation settings employed in FL studies, realistic on-device FL trainings exhibit complicated operational behavior patterns when it comes to energy consumption.
For example, edge devices with and without hardware accelerators (GPUs or TPUs) have different energy requirements for per sample throughout.
Furthermore, edge devices are usually executing workloads (i.e. baseloads) as such as inference or time-series data streaming which affects the energy per sample performance. 
These complex operational behavior patterns warrant a need for better energy management and energy-efficient FL. 
Client selection and scheduling are complex problems in FL, given the heterogeneous nature (hardware resources such as battery, compute, memory) of FL clients. 
Therefore, to enable better energy-efficient FL and carbon-efficient FL scheduling, metrics such as energy per sample, samples per second are necessary.
Current FL studies assume the energy availability budgets and do not consider energy per sample performance under the influence of power modes or baseloads of the FL clients~\cite{zhou2022joint,zaw2021energy,kim2021autofl,wiesner2023fedzero, guler2021framework}.

\section{Preliminary Insights}

To gain some preliminary insights into the variation of power-performance of FL clients, we evaluate an FL training under different baseloads and power modes of a Raspberry Pi.

For Raspbeery Pi, we evaluated three power modes: Performance, Powersave and Ondemand.  
To simulate a baseload on an FL device, we utilized unix command \emph{sysbench}\footnote{https://wiki.gentoo.org/wiki/Sysbench}. 
It provides flexibility to simulate baseload in-terms of number of threads/CPU cores.
For the FL training we utilized Flower framework \footnote{https://flower.dev}. We simulate a computer vision IoT training task using dataset CIFAR-10 \footnote{https://www.cs.toronto.edu/~kriz/cifar.html} and the computer vision model SqueezeNet which is light-weight and deemed to be suitable for edge computer visions applications. 
We assign higher kernel priority to baseload process to ensure the FL training doesn't affect the CPU time of baseload in co-running scenario. 
For the energy consumption measurement, we utilized WLAN power socket switch\footnote{https://www.delock.com/produkt/11826/merkmale.html}.
We report mean energy per sample and samples per second values for different power-modes and CPU core baseloads.  
\begin{equation}
  EPS = \frac{P_{total} - P_{BL}}{N}
  \label{eq_pps}
\end{equation}
\begin{align*}
  EPS & : \text{Energy per Sample} \\
  P_{\text{total}} & : \text{Total power consumption (FL and Baseload)} \\
  P_{\text{BL}} & : \text{Power consumption due to Baseload} \\
  N & : \text{Number of Samples}
  \end{align*}
Energy per sample values were calculated using Eq. \ref{eq_pps}.

  \begin{figure}[]
  
    \begin{subfigure}[b]{\linewidth}
      \centering
      \includegraphics[scale=.33]{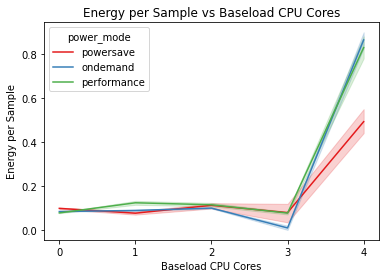}
      \caption{Energy Per Sample}
      \label{fig_energy_pi}
    \end{subfigure}
    
    \vspace{1em} 
    
    \begin{subfigure}[b]{\linewidth}
      \centering
      \includegraphics[scale=.33]{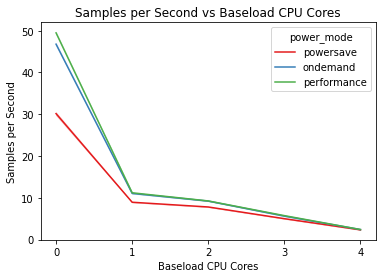}
      \caption{Samples Per Second}
      \label{fig_batches_pi}
    \end{subfigure}
    
    
    
    \caption{Power-Performance Characteristics for RPI}
  \end{figure}

Figure \ref{fig_energy_pi} illustrates the mean energy per sample and 95\% confidence intervals for each powermode, based on 10 repeated measurements.
We observe significant difference in energy per sample and samples per second values when there is no non-FL base load (0 baseload cores) compared to a scenario when non-FL baseload is executing and utilizing all CPU cores. 
We also observe that while samples per second (Figure \ref{fig_batches_pi}) doesn't vary significantly when non-FL baseload is co-running with FL, energy per sample values fluctuate for baseloads 3 and 4. 
For our experiments, ondemand mode with baseload cores 3 has an optimum energy usage when calculating same number of samples compared to other baseload and samples per second combinations.




\section{Conclusion and Future Work}
Recent research studies have focused on energy-efficient and carbon-efficient FL scheduling and client selection.
However, most of the research assumes simplistic energy consumption models for underlying FL clients. 
In this work, we showed that how energy per sample values under real-world scenarios such as different power modes and non-FL baseloads at CPU cores can vary and exhibit complex operational behavior patterns.

For future work, following open research questions and possibilities could be explored further,
\begin{itemize}
  \item How do current FL systems communicate FL clients' energy related information? 
  How to collect energy per sample, throughput per second and uncertainty related information at runtime?
  \item How can we predict the power-performance characteristics, what are the relevant metrics?
  With more data about real-world impact factors affecting energy footprint of edge devices, can we build predictive models for forecasting?
  \item How often do we need to measure before we can be certain? 
  Can we report the uncertainty to be used in scheduling? 
  FL trainings are usually executed multiple times due to data distribution drifts and hyperparameter search. 
  This repetitive FL training execution could be leveraged to collect more data about power-performance behavior patterns of FL clients.
  \item What's the impact of  hardware accelerated edge devices such as jetson nano on energy related metrics? 
  What are the energy efficiency opportunities in FL and non-FL co-running scenarios?
\end{itemize}



\bibliography{bibliography}
\bibliographystyle{ieeetr}

\end{document}